\documentstyle[12pt,epsf]{article}
\input{psfig}
\begin{document}
\noindent
{\small
Contribution to the Proceedings of ''RHIC'97 Summer Study''\\
July 7 - 16, 1997, Brookhaven National Lab.}\\[3mm]
\begin{center}
{\Large \bf
Estimates of dilepton spectra from open charm and\\[1mm]
bottom decays}\footnote{
Supported by BMBF grant 06DR829/1.}\\[1cm]
{\sc B. K\"ampfer$^a$, O.P. Pavlenko$^{a,b}$, K. Gallmeister$^a$}\\[1cm]
$^a$Research Center Rossendorf, PF 510119, 01314 Dresden, Germany\\
$^b$Institute for Theoretical Physics, 252143 Kiev, Ukraine
\end{center}

\vspace*{9mm}

{\small
The spectra of lepton pairs from correlated
open charm and bottom decays in ultrarelativistic heavy-ion collisions
are calculated. Our approach includes energy loss effects
of the fast heavy quarks in deconfined matter
which are determined by temperature and density of the expanding parton
medium. We find a noticeable suppression of the initial transverse momentum
spectrum of heavy quarks due to the energy loss
within the central rapidity covered by the PHENIX detector system.
As a result the dilepton yields from correlated decays of heavy quarks are
in the same order of magnitude as the Drell-Yan signal.
} 

\section{Introduction}

The production of high invariant mass dileptons in relativistic
heavy-ion collisions has attracted a great deal of interest in the recent years
from both the experimental and theoretical sides.
Dilepton measurements are planned
with the PHENIX and ALICE detector facilities at the
relativistic heavy-ion collider RHIC in Brookhaven and at the large hadron
collider LHC in CERN, respectively.
In despite of many theoretical attempts
the nature of the dominating source of dileptons is still unclear.
Among the candidates that can give
strong contributions in the large invariant mass continuum region near and
beyond the $J/\psi$ one can consider the following processes in which
dileptons are generated:
(i) hard initial quark - anti-quark collisions
give rise to the Drell-Yan yield,
(ii) semi-leptonic decays of heavy quarks like the charm and bottom,
which are also produced dominantly in hard initial parton interactions,
and
(iii)
the thermal dilepton radiation resulting in
interactions between secondary partons or even in a
locally thermalized quark-gluon plasma (QGP).
The yield from (iii) is expected to serve as a direct probe from
the QGP \cite{Shur3,Ruusk}.

It is commonly believed that, due to the large invariant mass $M$, at least
the Drell-Yan process and the initial heavy quark production in pp and
pA collisions are under reliable theoretical control by means of
perturbative QCD
and, therefore, can provide a reference for the thermal dilepton
signal from deconfined matter. At the same time all up to date calculations
of high-$M$ thermal dilepton rates from early parton matter contain
uncertainties due to the lack of knowledge of precise initial condition
parameters such as the energy density and quark-gluon phase space saturation.
Recent estimates
\cite{our_PL} within the mini-jet mechanism of quark-gluon matter formation
\cite{Eskola}, with initial conditions similar to the
self-screened parton cascade model \cite{E.M.W.}, point to a strong
competition between the Drell-Yan yield and the thermal signal in a wide
range of beam energies.

With respect to the wanted dilepton signal from the QGP
the dileptons from charm and bottom decays play a r\^ole of
substantial background \cite{Vogt2}.
The magnitude and the characteristic properties of such background is
presently matter of debate. As shown by Shuryak \cite{Shur1}
the energy loss effects of heavy quarks propagating through a medium
can drastically reduce the dilepton yield from charm and bottom decays
in the region of large invariant dilepton masses.
Further calculations \cite{Lin2}, including the PHENIX acceptance
at RHIC, also predict a very strong suppression of dileptons from heavy
quark decays, which then is partially below the Drell-Yan yield.
The studies in refs. \cite{Shur1,Lin2} are based on the
approximation of a constant rate of energy loss
$- dE/dx =$ 1 or 2 GeV/fm.
However, the value of $dE/dx$ is expected to depend
on the mean free path of heavy quarks
in the medium which is obviously governed a time dependent density.
As shown below, in particular within the gluon radiation mechanism
of energy loss \cite{Baier},
$dE/dx$ can depend on widely varying characteristic quantities
of the parton medium.

In the present contribution we study the dileptons resulting
from decays of correlated open charm and bottom
produced initially in hard collisions and undergoing an energy loss
which is determined by the temperature and the density evolution
of the deconfined medium.

\section{Initial hard production processes}

In the  framework of perturbative QCD the number of the heavy
quark - anti-quark ($Q \bar Q$) pairs, produced
initially with transverse momenta $p_{\perp 1} = -p_{\perp 2} =  p_\perp$
at rapidities $y_1$ and $y_2$, in
central AA collisions can be calculated from
\begin{equation} \label{eq.1}
dN_{Q \bar Q}= T_{AA}(0) \, {\cal K}_Q \, H(y_1,y_2,p_\perp) \,
dp_\perp^2 \, dy_1 \, dy_2,
\end{equation}
where $H(y_1,y_2,p_\perp)$ is the standard combination
of structure functions and elementary cross sections
(see for details \cite{our_PL,Vogt2})
\begin{eqnarray}
H(y_1,y_2,p_\perp)
& = &
x_1 \, x_2 \left\{ f_g(x_1,\hat Q^2) \, f_g(x_2,\hat Q^2)
\frac{d \hat \sigma_g^Q}{d \hat t} \right. \nonumber \\
& + &
\left. \sum_{q \bar q} \left[
f_q(x_1,\hat Q^2) \, f_{\bar q} (x_2,\hat Q^2) +
f_q(x_2,\hat Q^2) \, f_{\bar q} (x_1,\hat Q^2)\right]
\frac{d \hat \sigma_q^Q}{d \hat t} \right\},
\end{eqnarray}
where $f_i(x,\hat Q^2)$ with $i=g,q,\bar q$
are the parton structure functions,
$x_{1,2} = m_\perp \left(
\exp\{ \pm y_1 \} + \exp\{ \pm y_2 \}  \right) /\sqrt{s}$ and
$m_\perp = \sqrt{p_\perp^2 + m_Q^2}$.
As heavy quark masses we take $m_c =$ 1.5 GeV and $m_b =$ 4.5 GeV.
We employ throughout the present paper
the HERA supported structure function set MRS D'- from the PDFLIB
in CERN. The overlap function for central collisions is
$T_{AA}(0) = A^2/(\pi R_A^2)$
with  $R_A = 1.1 A^{1/3}$ fm and $A = 200$ in this paper.
We do not include shadowing effects of nuclear parton distributions in the
present paper since for heavy quark production
they are expected to be not very important
and can be considered separately.
Our calculation procedure is based on the lowest-order QCD
cross sections $d \hat \sigma_{q,g}^Q / d\hat t$
for the subprocesses $gg \to Q \bar Q$ and $q \bar q \to Q \bar Q$
with the simulation of higher order corrections by the corresponding
${\cal K}_Q$ factor.
Such a procedure reproduces within the needed
accuracy the more involved next-to-leading order calculations \cite{Vogt2}
for the heavy quark pair
distributions with respect to their invariant mass,
total pair rapidity and gap rapidity. We find the scale $\hat Q^2=4m_Q^2$
and ${\cal K}_Q=2$ as most appropriate.

For the Drell-Yan production process of leptons at rapidities $y_1$ and
$y_2$ and transverse momenta $p_{\perp 1} = - p_{\perp 2} = p_\perp$
we have
\begin{eqnarray} \label{eq.1l}
dN_{l \bar l}^{DY} & = & T_{AA}(0) \, {\cal K}_{DY} \, L(y_1,y_2,p_\perp) \,
dp_\perp^2 \, dy_1 \, dy_2, \\
L(y_1,y_2,p_\perp) & = & \sum_{q,\bar q} x_1 x_2
\left[
f_q(x_1,\hat Q^2) \, f_{\bar q} (x_2,\hat Q^2) +
f_q(x_2,\hat Q^2) \, f_{\bar q} (x_1,\hat Q^2) \right]
\frac{d \hat \sigma_q^{l \bar l}}{d \hat t},
\end{eqnarray}
with
$d \hat \sigma_q^{l \bar l} / d \hat t = \frac{\pi \alpha^2}{3}
\mbox{ch} (y_1 - y_2) / \left( 2 p_\perp^4
[1 + \mbox{ch}(y_1 - y_2)]^3 \right)$,
$x_{1,2} = p_\perp \left(
\exp\{ \pm y_1 \} + \exp\{ \pm y_2 \} \right)$ $/\sqrt{s}$,
$\alpha = 1/137$, $\hat Q^2 = x_1 x_2 s = M^2$
and ${\cal K}_{DY} =$ 1.1.

\section{Propagation of the heavy quarks through the parton medium
and energy loss}

Since the energy loss of heavy quarks depends strongly on the properties
of produced matter, which is propagated through,
one has to specify the initial conditions and the
space-time evolution of such a medium. To do this in a coherent
manner we employ for the mini-jet production (which dominates the parton
matter formation) the same lowest-order
approximation and parton structure functions
as for the initial heavy quark production.
Adding a suitably parametrized soft
component \cite{our_PL} we get the initial temperature
$T_i =$ 550 (1000) MeV, gluon fugacity $\lambda_i^g = 0.5$
and light quark fugacity $\lambda_i^q = \lambda_i^g / 5$
of the mini-jet plasma formed at initial time
$\tau_i = 0.2$ fm/c at RHIC (LHC).
The space-time evolution of the produced parton matter
after $\tau_i$ is governed by the longitudinal scaling-invariant expansion
accompanied by
quark and gluon chemical equilibration processes \cite{Biro,our_PRC}.
We take for definiteness full saturation, i.e.
$\lambda^g = \lambda^q = 1$ at deconfinement temperature $T_c = 170$ MeV.
\footnote{
Due to gluon multiplication this seems to be justified, while
the chemical equilibration of the quark component might be somewhat
overestimated by this ansatz.}
The actual time evolution
is determined by $e \propto \tau^{-4/3}$ for energy density and
$\lambda^{q,g} \propto \tau^2$ \cite{our_PRC}.

To model the energy loss effects of heavy quarks in expanding matter we assume
as usual in the Bjorken scaling scenario
that a heavy quark produced initially
at given rapidity follows the longitudinal flow with the same
rapidity. Therefore, with respect to the fluid's local rest frame
the heavy quark has essentially only a transverse momentum $p_\perp(\tau)$
depending on the proper local
time $\tau$ in accordance with the energy loss in transverse direction.

To calculate the evolution of the transverse momentum $p_\perp(\tau)$ of
quarks propagating the distance $r_\perp (\tau)$ in transverse direction
we adopt the results of Baier et al. \cite{Baier} for the
energy loss of a fast parton in a hot QCD medium
\begin{eqnarray}
\label{eq.21}
\frac{d p_\perp}{d \tau} & = & - \frac43
\frac{\alpha_s k_c}{\sqrt{L}} \, (p_\perp^2 + m_Q^2)^{1/4} \,
\ln \left( \frac{\sqrt{p_\perp^2 + m_Q^2}}{L \, k_c^2} \right),\\
\label{eq.22}
\frac{d r_\perp}{d \tau} & = & \frac{p_\perp}{\sqrt{p_\perp^2 + m_Q^2}},
\end{eqnarray}
where the parameter
$k_c^2 = 2 m_{th}^2(T) \equiv \frac 89 (\lambda_g +\frac 12 \lambda_q)
\pi \alpha_s T^2$
\cite{Biro,our_PRC} is used as an average of the
momentum transferred in heavy quark-parton scatterings,
and the strong coupling is described by $\alpha_s = 0.3$.
For the mean free path of gluons we take
$L \approx (\sigma_{gg} n_g)^{-1}$ with integrated cross section
$\sigma_{gg} \sim \alpha_s^2/k_c^2$ and gluon density
$n_g \sim \lambda_g T^3$,
resulting in $L^{-1} = 2.2 \, \alpha_s \, T$
\cite{Biro}.
The derivation of eqs.~(\ref{eq.21},\ref{eq.22}) relies on the
assumption that the
temperature changes smoothly enough in space-time so that the mean free path
of heavy quarks is less than the characteristic scale of the temperature
gradient. Such an approximation can be improved by considering
finite size effects \cite{Baier2}.
The r.h.s. of eq.~(\ref{eq.21}) is actually the rate of energy loss $dE/dx$
which is essentially $\propto (p_\perp^2 + m_Q^2)^{1/4}$.
Due to the rapid expansion of the matter this value
depends strongly on the time.
For instance, at RHIC conditions and initial momentum of a heavy quark of
5 GeV the stopping power $- dE/dx$ drops from 2.8 GeV/fm down to 0.5 GeV/fm
during the expansion of deconfined matter.

To get realistic spectra of charm and bottom quarks
after energy loss we use a Monte Carlo
simulation with a uniform distribution of the random initial position
and random orientation of $Q \bar Q$
pairs in the transverse plane.
We integrate then eqs.~(\ref{eq.21},\ref{eq.22}) together with the
above described time evolution of
$T(\tau)$ and $\lambda_{q,g}(\tau)$. Eq.~(\ref{eq.22}) is used to check
whether the considered heavy quark propagates still within
deconfined matter; if it leaves the deconfined medium it does not longer
experience the energy loss according to eq.~(\ref{eq.21}).
(The subsequent energy loss in a possible mixed and hadron phase might
be incorporated along the approach of \cite{Sv}.)
A considerable part of heavy quarks can not
escape the parton system thereby undergoing thermalization.
Following \cite{Lin2}
we consider a heavy quark to be thermalized if its transverse mass
$m_\perp$
during energy loss becomes less than the averaged transverse mass
of thermalized light partons at given temperature.
This part of heavy quarks we assume to be distributed as
$dN_Q / dm_\perp \propto m_\perp^2 \exp\{ -m_\perp / T_{\rm eff} \}$
with an effective temperature $T_{\rm eff} = 150$ MeV.

The results of our calculation for
transverse momentum spectra of charm and bottom quarks
at RHIC are shown in fig.~1.
We obtain a noticeable suppression of the initial transverse momentum
spectrum at large transverse momenta due to the energy loss both for charm and
bottom. The bump below 3 GeV is caused by the thermal redistribution.
In comparison with LHC conditions \cite{FZR-188}
the suppression at RHIC is found to be less pronounced due to the shorter
life time of deconfined matter and its lower initial temperature and density.
It should be stressed that such a change of the $p_\perp$ spectrum
of heavy quarks
in AA collisions, in comparison with pp or pA collisions,
is a measurable effect which for itself can help to
identify the creation of a hot and dense parton gas \cite{Tannenbaum}.

\section{Dilepton spectra from heavy quark decays}

Basing on the analysis of ref.\ \cite{Vogt2} we employ a $\delta$ like
fragmentation function for heavy quark fragmentation into D and B mesons.
The correlated lepton
pairs from $D \bar D$ and $B \bar B$ decays have been obtained from a suitable
Monte Carlo code which employs the single electron and positron
distributions as delivered by JETSET 7.4.
The average branching ratio of $D \to l + X$ is taken to be 12\%.
We consider here only the so-called primary leptons \cite{Lin2}
which are directly produced in the bottom decay
$b \to l \, X$ with the average branching ratio 10\%.
The secondary leptons $l'$ produced in further branchings like
$b \to c \, X \to l' Y$ have much smaller energy and do not contribute
to the large invariant mass region. Such simplification does not affect
significantly the dielectron yield at $M > 2$ GeV as we have checked by
comparison with ref.\ \cite{Vogt2}.
To get the dielectron spectra from charm and bottom
decays at RHIC we take into account the acceptance of the PHENIX detector:
the electron pseudo-rapidity is $|\eta_e| < 0.35$
with an additional cut on the electron transverse momentum
$p_\perp^e > 1$ GeV
to reduce the lepton background from light hadron decays.
Our results of heavy quark decays are depicted in fig.~2 for charm
(a) and bottom (b) for $M \ge 2$ GeV.
Clearly seen is that both the charm and bottom signals are suppressed
so that they compete with the Drell-Yan yield. Changing the kinematical
gates (i.e., narrowing rapidity or enlarging the minimum $p_\perp$)
does not help to suppress the charm and bottom signals below Drell-Yan.

To demonstrate the general tendency of the relative contribution from the
Drell-Yan process and heavy quark decays with the change of the collider
energy we plot in fig.~3 the results of our calculations for LHC initial
conditions including the ALICE detector acceptance for the
electrons, i.e. $|\eta_e| \le 0.9$.
As above we here also employ the gate $p_\perp^e > 1$ GeV.
One observes that the strong energy loss at LHC is not enough to
suppress sufficiently the signal from bottom decays, while the charm yield
is clearly below the Drell-Yan yield. For more details see \cite{FZR-188}.

\section{Summary}

In summary, taking into account the energy loss effects of heavy
quarks we calculate the spectra of electron pairs
from correlated open charm and bottom decays. We obtain
a measurable suppression of large transverse momenta
of heavy quarks due to the energy loss in the expanding deconfined
matter. Our calculations predict competing contributions
of charm and bottom decays and Drell-Yield
in the high invariant mass region of electron pairs which are planned
to be measured at central rapidity in the PHENIX detector.
Further studies, including a different fragmentation scheme and other
kinematical cuts, are in progress.

{\bf Acknowledgement:}
Stimulating discussions with participants of the RHIC'97 workshop
are gratefully acknowledged.


\newpage

\begin{figure}
\vskip -1cm
\centerline{{\psfig{file=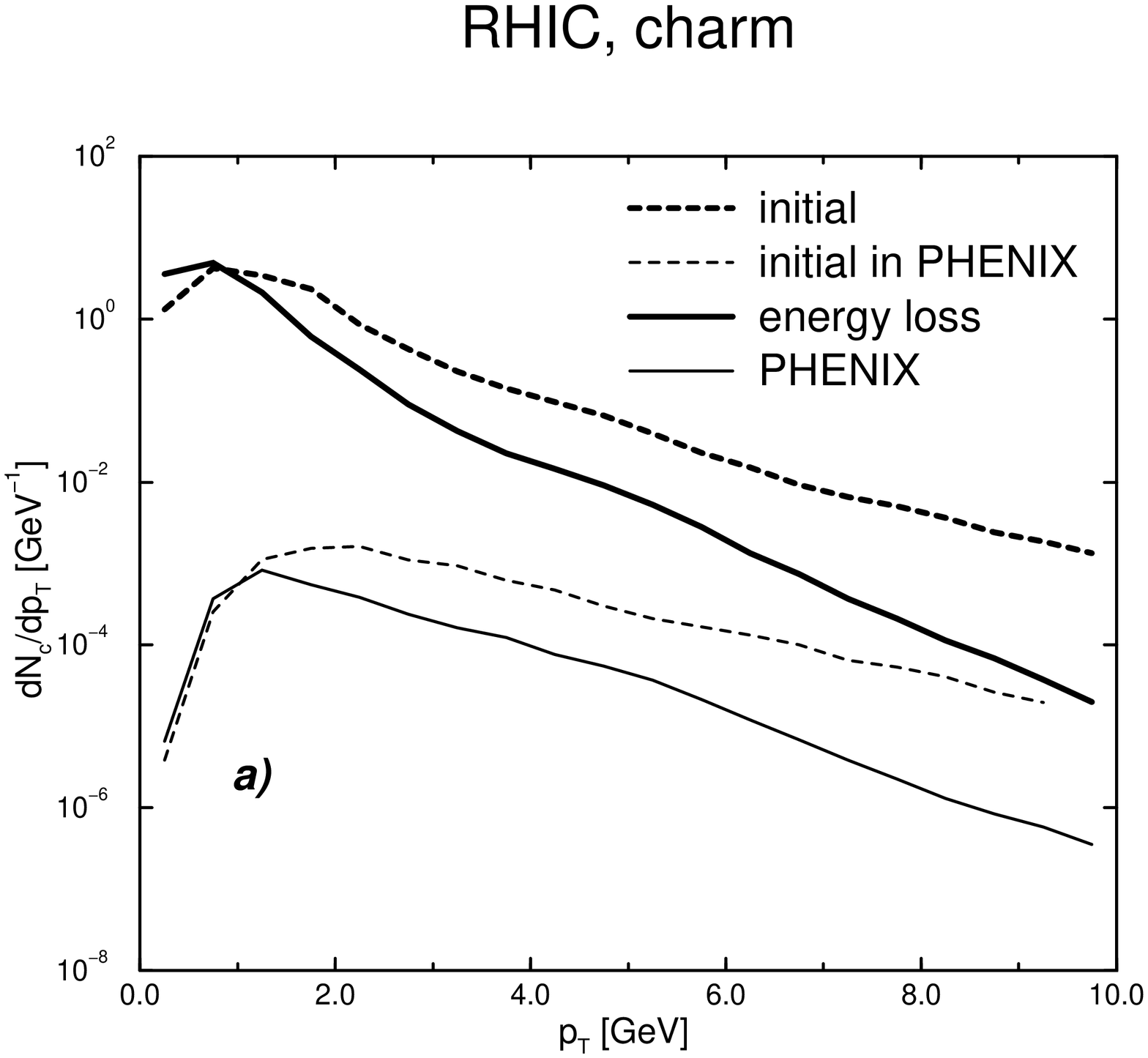,width=10cm,angle=-90}}}
\vskip 1cm
\centerline{{\psfig{file=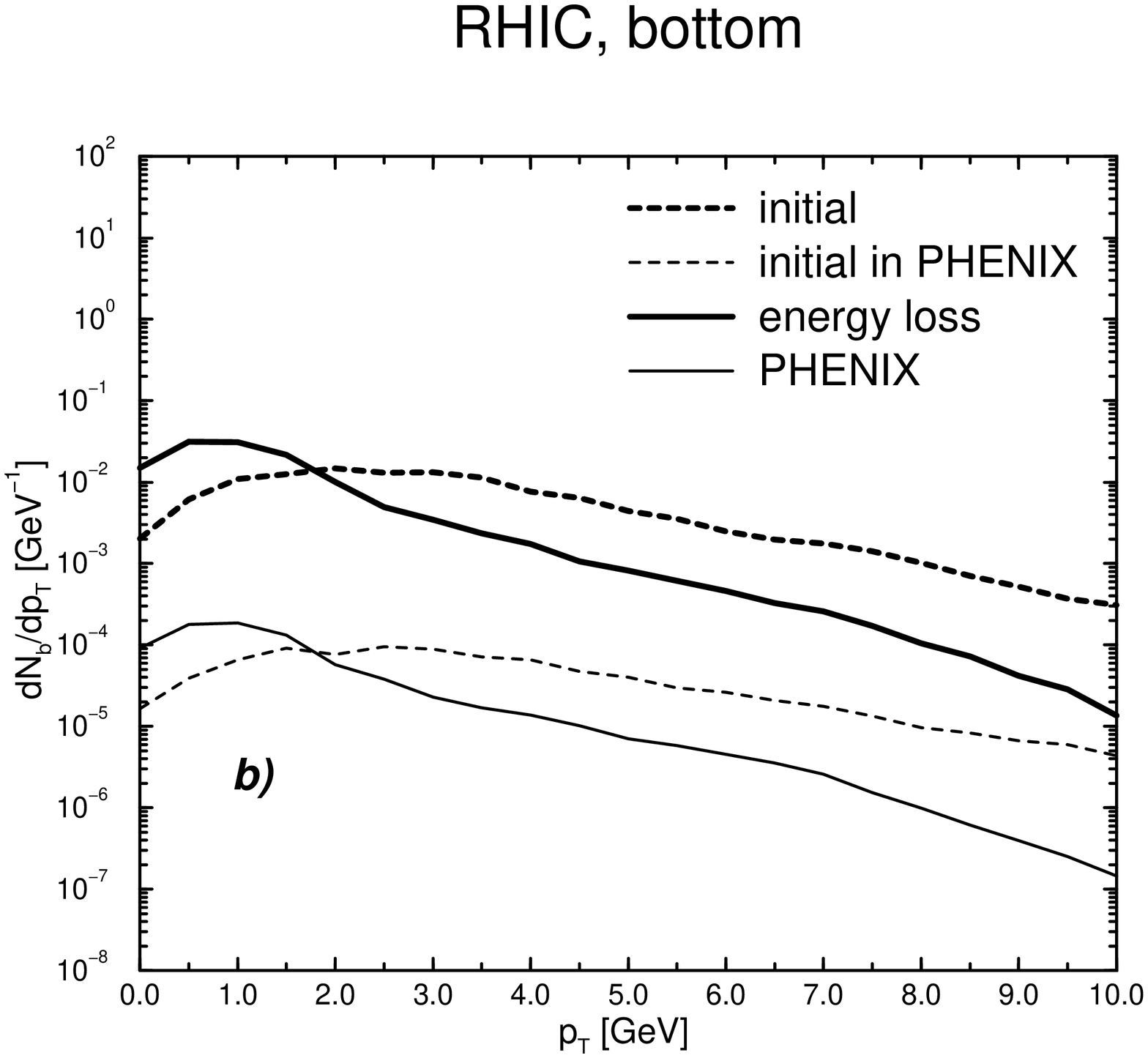,width=10cm,angle=-90}}}
\vskip .1cm
\caption{
The $p_\perp$ spectrum of heavy quarks ((a): charm, (b): bottom)
in central AA collisions at RHIC. The dashed curves depict
the initial production without energy loss,
while the solid lines show our result
with energy loss and $T_{\rm eff} = 150$ MeV for thermalized heavy quarks.
Fat curves are for full phase space,
while the light curves display these
heavy quark pairs which are detected in PHENIX through
their primary decay dielectrons.
}
\label{Fig.1}
\end{figure}

\newpage
\begin{figure}
\vskip -1cm
\centerline{{\psfig{file=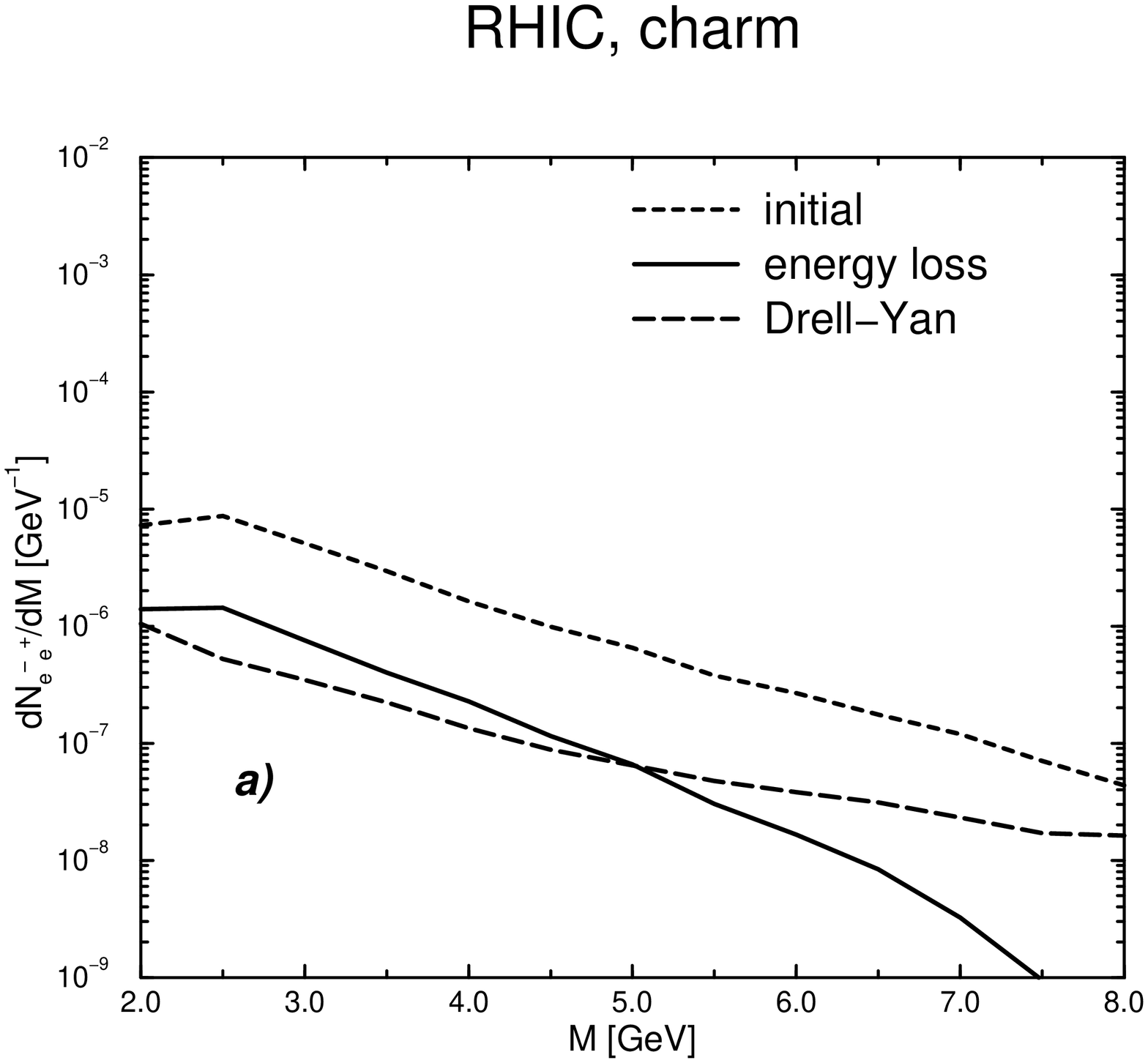,width=10cm,angle=-90}}}
\vskip 1cm
\centerline{{\psfig{file=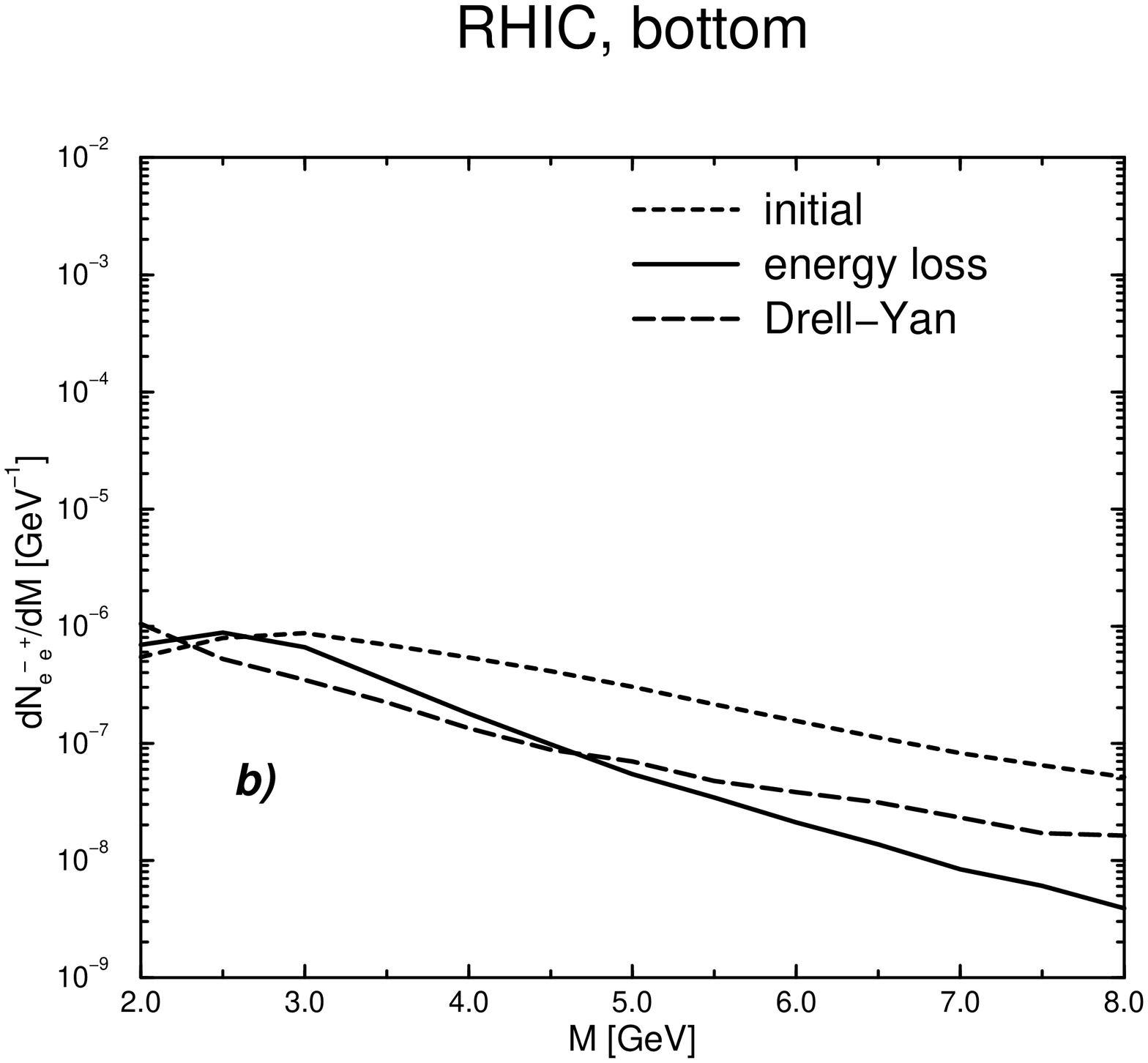,width=10cm,angle=-90}}}
\vskip .1cm
\caption{
The invariant mass spectrum of dielectrons from correlated charm (a) and
bottom (b) decays as well as Drell-Yan electron pairs
filtered throughout the PHENIX acceptance.
}
\label{Fig.2}
\end{figure}

\newpage

\begin{figure}

\vspace*{-6cm}

\centerline{{\psfig{file=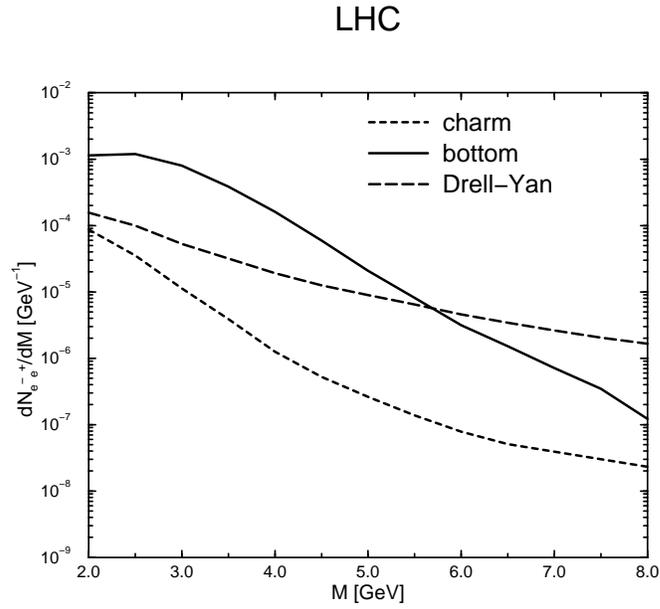,width=10cm,angle=-90}}}
\vskip .1cm
\caption{
The invariant mass spectrum of dielectrons from correlated charm
(short-dashed line)
and bottom (solid line) decays after enegy loss
as well as Drell-Yan electron pairs (long-dashed line)
filtered throughout the ALICE acceptance as described in text.
}
\label{Fig.3}
\end{figure}
\end{document}